\documentstyle[12pt,draft]{article}
\begin{document}
\begin{center}
\Large \bf
 Relativistic Effects in Heavy-Ion Collisions
at SIS Energies\footnote{Work supported by the GSI-Darmstadt
and the BMFT under contract 06T\"U736.}
\vspace*{1cm}\\
\normalsize \bf\large
Rajeev K. Puri
, E. Lehmann, Amand Faessler\footnote{e-mail: Faessler@
mailserv.zdv.uni-tuebingen.de} and S.W. Huang\\
\normalsize\it
Institut f\"ur Theoretische Physik der Universit\"at T\"ubingen,\\
 Auf der Morgenstelle 14, D 72076, T\"ubingen, Germany.
\vspace*{1cm}
\\
\today
\vspace*{1cm}
\\
{\bf Abstract:}\\
\end{center}
The covariant and non-covariant Quantum Molecular Dynamics
models are
applied to investigate possible relativistic effects  in heavy
ion collisions at SIS
energies. These relativistic effects which arise due to
the full covariant treatment of the dynamics are studied at
bombarding energies E$_{lab.}$ = 50, 250, 500, 750, 1000, 1250,
1500, 1750 and 2000 MeV/nucl.. A wide
 range of
the impact parameter from b = 0 fm to b = 10 fm
is also considered. In the present study, five systems
$^{12}$C-$^{12}$C, $^{16}$O-$^{16}$O, $^{20}$Ne-$^{20}$Ne,
 $^{28}$Si-$^{28}$Si and
$^{40}$Ca-$^{40}$Ca are investigated. The full covariant treatment
at low energies shows quite good agreement with the corresponding
non-covariant
approach whereas at higher energies
it shows  less stopping  and hence less thermal equilibrium
as compared to the non-covariant approach. The
 collisions dynamics is less affected. The
density using RQMD rises and  drops
 faster than with QMD. The relativistic effects show some
influence on the
resonance matter production. Overall, the relativistic effects at SIS
energies ($\leq$ 2000 MeV/nucl.) are less significant.


\normalsize
\newpage
\baselineskip 24pt
{\large \bf 1. Introduction :  }\\
\vspace*{0.3cm}\\
In the beginning of the last decade it was thought that by
comparing the predictions of the theoretical models with existing
experimental
data, one may be able to conclude something about the {\bf
E}quation {\bf o}f {\bf S}tate (EOS). Unfortunately, it turned
out that the EOS is drastically  affected by the ingredients of
these models.
Therefore, up to now the question of the
equation of state is still open.
 In the intermediate
energy region [ $\leq$ 2000 MeV/nucl.], both main inputs of the nuclear
 dynamics, namely,
the mean field ( or mutual two- and three body interactions) and
the Nucleon-Nucleon cross-section are found to
influence the nuclear dynamics to a
larger extent [1-6].

Apart of the choice of possible different  interactions and cross-sections, one
has to remember that  even at these intermediate energies,
the velocities of particles are not at all negligible as compared
to the velocity of light. For example, at a bombarding
energy of 1 GeV/nucl., the boost velocity is about 85$\%$ of
the speed of light. Thus, even at these energies,
the Lorentz-invariance of the theory has to be respected. Once the
bombarding energy of a nucleon becomes comparable with its rest mass,
the relativistic effects are expected to influence the dynamics.

To study the nuclear dynamics at intermediate energies where no
local or global equilibrium is reached, one needs a transport
theory which is not based on the assumption  of local or global
equilibrium. In this energy domain, the
Boltzmann-Uehling-Uhlenbeck (BUU) and Quantum Molecular Dynamics
(QMD) models with their relativistic versions i.e. RBUU and RQMD
are used with big success [1-23].

Recently, Faessler and collaborators have reported their new covariant
generalization of the QMD model [i.e. Relativistic Quantum
Molecular Dynamics (RQMD)]\cite{lehmann1994}.
 The first RQMD was made by the
Frankfurt group \cite{sorge1989}.
 One of the main advantages of our
  RQMD is that we have shown for practically the first time
 that this numerical implementation
of a covariant theory gives in the non-relativistic limit the same result(s)
as that of the corresponding non-covariant approach.
 At higher energies, clear relativistic effects in the flow were
  observed. One should note that the  transverse flow
 is a quite sensitive quantity. Therefore, to study
relativistic effects, one has to look for quantities
which are not very much sensitive to  model inputs or even
to different EOS's. Thus, quantities like resonance production,
density, rapidity distribution are better candidate for
checking  the validity of relativistic effects at SIS energies.
Hence, in this paper we concentrate fully on these quantities
and search for relativistic effects.

We would like to remark here that the nature of  relativistic
effects can show up  very differently because there exits several origins
of these  effects. Some of them stem from the relativistic
kinematics, some from relativistic forces, retardation or meson
 radiation effects
etc. Parts of the relativistic kinematics are trivial and are
dealt even in the non-covariant models by using the relativistic
energy-momentum relations. But the dynamics in these models is
still treated in a non-covariant way and hence it breaks the
Lorentz-invariance of the theory. The relativistic effects which we
are going to discuss are the one which are originating due to the
full covariant treatment of the dynamics.

The paper is organized as follows: Section 2 deals with a brief
introduction to the formalisms used in  QMD and RQMD. The results
 are presented in section 3   and finally we
summarize our findings in section 4.
\\
\vspace*{0.2cm}
\\
{\large \bf 2.  The Formalism:}\\


The detailed formalism and the numerical realization of QMD and
RQMD are given in refs. \cite{aichelin1991} and \cite{lehmann1994}
 , \cite{sorge1989}, \cite{maruyama1991a}, respectively.
Here we discuss briefly the main important points of the
formalism used in the QMD and RQMD:\\

{\bf 2.1 Quantum Molecular Dynamics [QMD]: }

In QMD, the  nuclei under consideration are
 chosen by a procedure which is based
on the random choice of the coordinate and momentum space. This
is done with the help of a standard Monte-Carlo
procedure.  The nucleons are distributed in a sphere of radius
R = 1.14 A$^{1/3}$ which is consistent with the liquid drop
model. If the centers of the Gaussians of two nucleons are
closer than a distance R$_{min}$ = 1.5 fm, the choice of this
coordinate is rejected and other coordinates are chosen.
 The momenta of the nucleons are chosen randomly between
zero and the local Fermi momentum. The
successfully initialized nuclei are boosted towards each other
with proper center-of-mass velocity using relativistic
kinematics \cite{aichelin1991}.

In our approach each nucleon is  described
 by a Gaussian wave packet with a
width $\sqrt{L}$ centered around the mean position $\vec r_i$(t) and the
mean momentum $\vec p_i$(t):
\begin{equation}
\psi_i(\vec r,\vec p, t) = \frac{1}{(2\pi L)^{3/4}}
              ~~\mbox{exp}~\left\{-\frac{(\vec r - \vec r_i(t))^2}{4L}
                          +i \vec p_i(t) \cdot \vec r \right\},
\end{equation}
with L = 1.08 fm$^2$. This choice corresponds to a root mean square radius of
the nucleon wave-packet of  1.8 fm. As the width of the Gaussians
 is kept fixed, the  centeriods of the wave
-packets are propagated using the classical equation of motion.
\begin{equation}
\frac{d\vec r_i}{dt} = \frac{\partial H}{\partial \vec p_i},
\end{equation}
\begin{equation}
\frac{d\vec p_i}{dt} = -\frac{\partial H}{\partial \vec r_i},
\end{equation}
where the Hamiltonian is given by the classical N-body Hamiltonian
\begin{equation}
H=\sum_i\frac{\vec p_i^{\,2}}{2m_i} + V,
\end{equation}
with V as the potential, which in present study, is simple Skyrme
force
\begin{equation}
V=\sum_{i=1}^N\left[\frac{\alpha}{2}\left\{\sum_{j\neq i}
           \frac{\tilde\rho_{ij}}{\rho_0}\right\} +\frac{\beta}{\sigma+1}
            \left\{
           \sum_{j\neq i}\frac{\tilde\rho_{ij}}{\rho_0}\right\}^
           \sigma\right].
\end{equation}
Here  $\tilde \rho_{ij}$ is the interaction density which is given by
\begin{equation}
  \tilde\rho_{ij} =
                \int \rho_i(\vec r(t)) \rho_j(\vec r(t)) d \vec r
                = \frac{1}{(4\pi L)^{3/2}}~\mbox{exp}~
                  \left[- \frac{(\vec r_i -\vec r_j)^2}{4L} \right].
\end{equation}
The coefficients $ \alpha, \beta$ and $\sigma$  appearing in eq. (5) are
determined by the condition that the bulk properties of infinite
nuclear matter has to be reproduced. Different sets of
parameters lead to different  incompressibilities K which generate different
EOS's. Usually two  incompressibilities K are chosen: (i) K= 200 MeV
 corresponds to a soft EOS and
(ii) K = 380 MeV corresponds to a hard EOS \cite{aichelin1991}.

During the propagation, two nucleons are assumed to collide if
they come closer than a distance $A_{min}$=$\sqrt{\sigma(\sqrt{s})/\pi}$ where
$\sigma(\sqrt{s})$ is the total cross-section depending on the
invariant mass  $\sqrt{s}$.
 Whenever a collision
occurs, the phase space around the final state of the two scatterers
 is checked. From the overlap one is able to check the
probability whether a collision  is Pauli blocked or not.
 Here  we use the cross-sections
parametrized by Cugnon \cite{cugnon1981}.
 In these new cross-sections both elastic and
inelastic  channels are considered and different isospins
of baryons are also taken into account. Some  of the processes
 considered here are of the type:
 $NN \rightarrow NN$, $N\Delta \rightarrow N\Delta, NN \leftrightarrow
N\Delta,
\Delta\Delta \rightarrow \Delta\Delta$ etc. In the following, we
demonstrate the main points of the covariant QMD i.e of RQMD.
\\

\bf 2.2  Relativistic Quantum Molecular Dynamics [RQMD]:
\normalsize

The RQMD model describes the propagation of all kinds of
baryons and mesons in a Lorentz-invariant fashion. The
Hamiltonian for an N-particle system is expressed in terms of 8N
variables ( 4N position coordinates $q_{i\mu}$ and 4N momentum
coordinates $p_{i\mu}$).
This means that here each particle carries
its own energy and time. Since the physical events are
described as  world lines in a 6N dimensional phase-space,
extra 2N-1 degrees of freedom have to be eliminated and a global
evolution parameter $\tau$ has to be defined. This can be
achieved with the help of 2N constraints. In our approach,
the first N  constraints are chosen as
Poincar$\acute{e}$ invariant mass-on shell constraints
\cite{lehmann1994}, \cite{sorge1989}.
\begin{equation}
        \xi_i
= p_i^{\mu}p_{i\mu} - m_i^2 -\tilde V_i =0 \qquad ; \qquad i=1,...,N.
\end{equation}
This choice of Poincar$\acute{e}$ invariant constraints requires that
the potential part $\tilde V_i$ should be a Lorentz scalar and therefore
 function of Lorentz scalars only. Since in the RQMD, a
system with mutual two- and three-body interactions
 (like in  QMD)  has to be defined,  $\tilde
V_i$ should be given by the  sum of these two-body interactions.  Further,
as we want to look for   relativistic
effects in the dynamics,  we have to generalize the non-relativistic Skyrme
force in such a way that the force is  covariant
 and  reduces also to the  usual Skyrme force in the
non-relativistic limit. This
can be done as \cite{lehmann1994}
\begin{equation}
\tilde V_i = \sum_{j\neq i}^N\tilde V_{ij}(q^2_{Tij}).
\end{equation}
This shows  that the two-body interactions depend only on the
Lorentz invariant squared transverse distance
\begin{equation}
q^2_{Tij} = q^2_{ij} - \frac{(q^{\mu}_{ij}p_{ij\mu})^2}{p_{ij}^{\,\,2}},
\end{equation}
with $q^{\mu}_{ij}=q^{\mu}_i-q^{\mu}_j$ being the simple four
dimensional distance and $p^{\mu}_{ij}=p^{\mu}_i+p^{\mu}_j$ the
sum of the momenta of the two interacting particles $i$ and $j$.


The next set of constraints ( which fix the relative times
of all particles) should be chosen in such a way that
 these constraints must respect the principle of
 causality and N-1 of these constraints should be Poincar$\acute{e}$
invariant so that the world line invariance can also be
fulfilled.  Another feature which these constraints has to fulfill
 is the cluster separability. This means that the system can
be divided into single particles or clusters as soon as their
Minkowski distances are space-like.
 Furthermore,
 a global evolution parameter should also be defined.
These features can be fulfilled by choosing the following set of
time constraints:
\begin{equation}
\chi_{i} = \sum_{j(\neq i)}
 \frac{1}{q^2_{ij}/L_C}\mbox{exp}(q^2_{ij}/L_C)
{}~p^{\mu}_{ij}q_{ij\mu} = 0 \qquad ;
\qquad i=1,...,N-1,
\end{equation}
\begin{equation}
\chi_{2N} = \hat{P}^{\mu}Q_{\mu}-\tau = 0.
\end{equation}
with $\hat{P}^{\mu}=P^{\mu}/\sqrt{P^2}$, $P^{\mu}=\sum_i p^{\mu}_i$,
$Q^{\mu}=\frac{1}{N}\sum_i q^{\mu}_i$.

These time fixations take care that the time coordinates of
 interacting particles are not too much dispersed in the
center of mass system of two particles.
 The Hamiltonian is a linear combination of the
Poincar$\acute{e}$ invariant constraints:
\begin{equation}
     H = \sum_{i=1}^{2N-1}\lambda_i \Psi_i,
\end{equation}
with
\begin{equation}
\Psi_i = \left\{
 \begin{array}{c c l} \xi_i & ; & i \leq N \\
                      \chi_{i-N} & ; & N < i \leq 2N-1. \\
\end{array}
\right.
\end{equation}
This Hamiltonian then generates the
equations of motion
\begin{equation}
\frac{dq_i^{\mu}}{d\tau} = [H, q_i^{\mu}],
\end{equation}
\begin{equation}
\frac{dp_i^{\mu}}{d\tau} = [H, p_i^{\mu}].
\end{equation}
Here square brackets represent the Poisson brackets.
The unknown Lagrange
multipliers $\lambda_i$ in eq. (12) are determined by the condition that all
constraints must be fulfilled for all times during the
simulations. These equations of motion are used to propagate the
baryon during the reaction.

The propagation and the "soft interaction" between baryons
is combined with the quantum effects like
 stochastic scattering and the Pauli-blocking etc..
 In RQMD, the collision part is treated in a
covariant fashion. Therefore all quantities which determine
the collision must be Lorentz invariant. In RQMD
 two baryons are allowed to collide if their distance
$\sqrt{-q_{Tij}^2}  \leq \sqrt{\sigma(\sqrt{s})/\pi}$
 where $q_{Tij}^2$ is the Lorentz-invariant squared transversal
distance (eq.9) and $\sigma(\sqrt{s})$ is the cross-section depending
on the available invariant mass $\sqrt{s}$.


\vspace*{1cm}

{\large \bf 3. Results and Discussion:}
\\

All results presented here are calculated using a new
simulation package which integrates the RQMD and QMD approaches
under one shell and hence this code has been named as
UNISCO which stands for {\bf UNI}fied
{\bf S}imulation
{\bf CO}de \cite{lehmann1994}.
 In this paper, five systems
$^{12}$C-$^{12}$C, $^{16}$O-$^{16}$O,
$^{20}$Ne-$^{20}$Ne, $^{28}$Si-$^{28}$Si and
$^{40}$Ca-$^{40}$Ca are considered.
For a detailed investigation
of the relativistic effects, we simulate these systems at  bombarding
energies $E_{lab}$ = 50, 250, 500, 750, 1000, 1250, 1500, 1750
and 2000 MeV/nucl.. In addition, a wide range of the impact
parameters between b = 0 ( central collisions) and b = 10 fm (
peripheral collisions)  is also considered.

We start with the time evolution of nuclear density. Here the density
is calculated in a sphere with a radius of 2 fm.
The center of this sphere is located at the point where the two
nuclei touch each other in their center of mass system.
Fig. 1 shows the time evolution of the maximum density reached
 in this sphere. Here a semi-central collision of
$^{40}$Ca-$^{40}$Ca is considered at a bombarding energy 1.5 GeV/nucl.
 with both hard and soft EOS's.  It is quite
interesting to note that RQMD and QMD simulations show a quite
different evolution of the maximum density.
Compared to  QMD, the coordinate space
in RQMD is
Lorentz-contracted and this leads to more repulsion in the RQMD
 and the  density decreases faster.
  This rapid decrease in the density shows that the particles
are kicked out from the compressed zone and one should get
less thermalization using RQMD than
 QMD. Though the maximal values are not much different,
the full shape of evolution of the density is quite different.


 What makes this difference? Is it only a Lorentz
contraction of the initial phase space distribution? To answer these
questions we give the main differences between RQMD and QMD which are
clearly visible in addition to the covariant feature of the RQMD model:
(i)  In RQMD, we have  an initial Lorentz-contracted distribution in coordinate
space and an elongated distribution in the momentum space. We will come
to this point later when QMD simulations with this initial
Lorentz-contracted distribution will be presented.
 (ii)  In RQMD, a multi-time formalism is used. In other words,
in RQMD all baryons carry their own time coordinates.
 (iii) RQMD has a full covariant treatment of the collision part
which also includes Pauli-blocking covariantly \cite{sorge1989}.
(iv) In RQMD, the mean field is a
Lorentz-scalar whereas in  QMD it is a zero component of the Lorentz
vector. Due to the covariant feature, the interactions in RQMD are
defined as a function of the distance between the particles in the rest
frame of their common center of mass. Therefore, in a moving frame ( e.g.
in the reference frame corresponding to the center of mass of
the two nuclei) these interactions are  not
 spherical but are Lorentz contracted in the direction of the motion
 of the two particles. Therefore, the strength of the interaction depends
 strongly on the direction of the center of mass motion of the two
 nucleons in the rest frame of the two nuclei. When the initial
phase space
distribution is Lorentz contracted then, naturally, the density of a
 fast moving nucleus is increased in the CM system. If
  one includes Lorentz contraction in normal QMD then one finds
that it can lead to a tremendous enhancement in the
transverse flow \cite{lehmann1994}.
 When we use the
feature in covariant RQMD,  this artificial repulsion due
to the initial contraction of the phase space is partially
counterbalanced.

To understand these effects we follow in fig.2 the time evolution
  of the collisions rate
for  the same reaction as in fig.1 but  at an
impact
parameter of 1 fm. One sees that the collision
evolution reflects the density behaviour. The QMD simulations show
the first collisions  around 4 fm/c whereas RQMD
 only around 8 fm/c. The collision rate in RQMD falls more rapidly than
in QMD. We further note that this behaviour is the same at all
higher energies and for all impact parameters.
 In fig.3, we show the time evolution of the  number of
collisions for the systems
$^{12}$C-$^{12}$C, $^{16}$O-$^{16}$O, $^{20}$Ne-$^{20}$Ne,
 $^{28}$Si-$^{28}$Si
 at a bombarding energy of 1.5 GeV/nucl. and
 for an impact parameter = 0.25 b$^{max}$ ( $b^{max}$ =
 the radius of target + the radius of projectile).
 Here the hard EOS is used.
 It is interesting that all reactions show a similar behaviour for the
time evolution of the collision rate. The size of the
relativistic effects is similar in all reactions.

A further decomposition of the collision rate into
elastic and inelastic
channels is shown for the case of $^{40}$Ca-$^{40}$Ca in fig.4.
The elastic
 channel in RQMD seems
to dominate over the QMD  whereas, in the case of inelastic
channel, a careful
look shows that the situation is not clear.
The inelastic channel which contains the formation of resonance
matter will be considered later on.  In order to look how much collisions
one can get when one includes the Lorentz contraction
 in coordinate space in a normal QMD and keeps
  the interaction still spherical, we
show in fig. 5 the reaction  $^{40}$Ca-$^{40}$Ca at 1.5 GeV/nucl.
 In this figure, the
time evolution of the total number of collisions per nucleon
is shown using
the usual RQMD and QMD
and a special version of QMD where
 the initial phase space
is Lorentz boosted (it is labeled as QMD(cont.)). It is interesting
 to see that RQMD shows more collisions than the
normal QMD but this collision number is less than
what we get using QMD(cont.).
This justify our earlier claim that the covariant
treatment of the interactions counterbalances partially
 the initial contraction.
 In fig. 6, we switch
off the self-consistent field in RQMD, QMD and QMD(cont.) i.e.
we use a cascade model. One sees that
now  RQMD  shows  more collisions than  QMD and
QMD(cont.). In a cascade-mode, all particles are
 allowed to move freely and
hence there is no more counterbalancing force. This shows that the
covariant treatment of the interaction keeps the nuclei not only
stable, but  counterbalances also the artificial collisions
which can happen due to the higher contracted  density.

One of the important quantum feature in
 all semi-classical models like BUU /QMD /RBUU /RQMD
 etc. is the inclusion  of the Pauli-principle. Therefore, in fig. 7,
we compare the percentage of the collisions which are blocked due
to the lack of free phase space using RQMD and QMD. In this figure,
the ratio of the Pauli-blocked collisions to all attempted collisions
 is shown for both the soft and the hard EOS's. The general feature i.e.
the decrease in the Pauli-blocked collisions with increase of the bombarding
energy is well reproduced using QMD and RQMD. Remember that in
RQMD, a
Lorentz-invariant Pauli-blocking procedure is implemented.
  It is clear from fig. 7 that at 50 MeV/nucl.
RQMD and QMD show good agreement as expected whereas at higher energies
( up to 2 GeV/nucl.), RQMD  shows more Pauli-blocking
 than QMD. This result which is valid
for all higher energies and masses shows that the covariant treatment
 of the theory results in more blocked collisions.

In a very recent paper, the name {\it resonance matter} has been
put forward \cite{ehehalt1993}, \cite{mosel1993} \cite{mina1994}.
 The name {\it resonance matter} is based on the fact that
 especially in central collisions, an appreciable
portion of nuclear matter is converted into excited resonances.
This resonance matter contains the $\Delta$'s and higher resonances.
This idea is , however, questioned recently by Frankfurt group
 \cite{bass1994}.
These studies of resonance matter are carried out with non-
covariant BUU and QMD. Thus, it is important to look for the effect
of a covariant treatment on resonance production.  This is done in
fig.8. Here the number of delta's which are obtained at the final
stage of the reaction (i.e. at 60 fm/c) are plotted as a function
of the impact parameter. We note that in central collisions, about
30 $\%$
nuclear matter is converted into delta matter.
We also note that up to semi-central collisions, the
relativistic effects show some reduction in the number of delta's.
This reduction in  the number of deltas using a covariant
theory can be
 important for the discussion of resonance matter at
relativistic energies.

Fig. 9 shows the mass dependence of the relativistic
effect on the delta production. Here five systems
$^{12}$C-$^{12}$C, $^{16}$O-$^{16}$O, $^{20}$Ne-$^{20}$Ne,
 $^{28}$Si-$^{28}$Si, $^{40}$Ca-$^{40}$Ca are considered.
The straight lines in this figure are the extrapolation of the delta
population per nucleon found in $^{40}$Ca-$^{40}$Ca to the  whole
mass region. In other words, e.g. in case of QMD (hard EOS),
we see that the delta population for $^{40}$Ca-$^{40}$Ca is 23
$\%$ ( about 18 deltas for 80 baryons).
 Therefore the straight line indicates the delta population
of 23 $\%$ for all mass region.  It is interesting to note that
the percentage of the
resonance matter is nearly independent of the mass of the
colliding nuclei. This result is in agreement with the finding of Ref.
\cite{ehehalt1993}. Further, it is also clear that the
relativistic effects in delta production are similar for all
masses considered here.
  In case of the  $^{40}$Ca-$^{40}$Ca collision, RQMD
simulations show about 12 $\%$ ( ($\Delta(RQMD)-\Delta(QMD)
)/\Delta(QMD)$ x 100 ) reduction
 in the delta population. This reduction can have some influence
for the production of kaons and other particles created
through mainly resonances \cite{huang1994}, \cite{spieles1993},
\cite{teis1992},
\cite{batko1994}.
 The influence of the bombarding energy on
relativistic effects in the delta production is shown in fig.10
 for the semi-central collision of
$^{40}$Ca-$^{40}$Ca.  We note that
the relativistic effects increase with the increase in  the
bombarding energy. For all energies,  one can
see an unique behaviour i.e. the reduction in the resonance matter
when one uses a covariant formalism.

It is also  interesting  to look for the thermalization in heavy
ion collisions. To study this, we analyse the rapidity
distribution which is a measure of the stopping of nuclear matter in
heavy ion collisions. The rapidity distribution  is defined as:
\begin{equation}
Y_i  =  \frac{1}{2} ~ \mbox{ln}~ \frac{E(i) + p_z(i)}{
       E(i) - p_z(i)},
\end{equation}
where E($i$) and $p_z(i)$ are the energy and the longitudinal momentum
of the ith particle. For full equilibrium, one should get  a
Gaussian shape peaked at mid-rapidity. In fig. 11, we
show the rapidity distribution at an impact parameter of 2 fm
using the hard  EOS. To establish the
relativistic effects over a wide range of energy,
 we show here the rapidity distribution
at four different incident energies i.e. at 50 MeV/nucl., 500 MeV/nucl.,
1 GeV/nucl. and 1.5 GeV/nucl.. We assumed that the final
rapidity distribution is reached  for 50 MeV/nucl. after 100 fm/c
 and  for higher energies after 60 fm/c. The
Rapidity distribution at 50 MeV/nucl. is the same for RQMD and QMD.
  At all higher energies
( from 500 MeV/nucl. to 1.5 GeV/nucl.),
the simulations using a covariant approach show less stopping than
the non-covariant. This is true for both the hard and the soft EOS's.
 One should note that  different forces and
in-medium effects modify the  transverse momentum and other quantities, but
the rapidity distribution is not affected
  \cite{khoa1992b}, \cite{jaenicke1992a},
\cite{puri1992a}, \cite{puri1994} \cite{aichelin1991},\cite{berenguer1992}.
 This reduction of the stopping power using RQMD was earlier predicted when the
time evolution of the density was discussed ( see fig.1). In this
figure 1 one sees a rapid decrease of the density in RQMD which shows
that the particles are stopped less in the hot and dense zone.
 These  results are in  nice
agreement with the available calculations of the Frankfurt group
\cite{hartnack1993a}.

In fig. 12, we show the
final rapidity distributions for the reaction  $^{40}$Ca-$^{40}$Ca
at the bombarding energy of 1.5 GeV
using  the hard EOS. Here we take
two extreme cases of the impact parameters i.e. a central
collision b = 0 fm  and a peripheral collision b = 6 fm.
 It is clear that in central
reactions,  the collision rate is very high and thus we
obtain a complete stopping. As one goes to semi-central collisions
(see fig. 11), the stopping starts to decrease
 and for peripheral collisions (fig.12), one sees nearly no stopping
 and hence one can still see two peaks at the target and projectile rapidities,
respectively.
Although the initial distributions in fig. 14 (at 6 fm/c)
are higher for QMD than for RQMD, the final rapidity
 distribution at impact parameter b=6 fm in
fig. 12 shows the opposite. It is also
clear that at all impact
parameters one has less stopping using a  full
covariant approach.
To see this further the rapidity distribution
 of $^{40}$Ca-$^{40}$Ca at 1 GeV/nucl.
is shown in fig. 13 calculated in a pure cascade-mode and in a pure
Vlasov-mode. The cascade-mode is obtained by switching off the
mutual "soft interactions" (self consistent field) between baryons and the
Vlasov-mode is obtained by switching off all collisions i.e.
we assume that all collisions are Pauli-blocked in a Vlasov-mode.
The cascade simulations  show the interesting
result that relativistic effects vanish in a cascade mode.
 Further,
when one compares  the 1 GeV/nucl. simulations using RQMD and QMD,
in fig.11 with
the cascade-mode in present fig., one sees that
the absence of the mean field produces
far more stopping.  The cascade mode
does not have any kind of repulsion even when nucleons are in
the hot and compressed central zone and thus it results in more stopping.
Due to the lack of any collisions, the Vlasov-mode shows no stopping.
Here one can  see  peaks at target and projectile rapidities.


In fig. 14, we follow the full time evolution of a
rapidity distribution
for a central collision for the same reaction, energy and impact
parameter as in fig.12.
 Here we show the result using QMD and RQMD
simulations at time t = 6 fm/c, 9 fm/c, 12 fm/c and 18 fm/c.
After 18 fm/c, the reaction at this high energy is practically
in its asymptotic state [see fig.1].
We also note that the soft and the hard EOS's show similar behaviour at all
times. At 6 fm/c, no collisions have occurred and hence it reflects the
situation of an initial rapidity distribution . One sees
that due to the Lorentz-
elongation of the momentum distribution, the RQMD simulations have
less high peaks at target and projectile rapidities but the
Gaussians are broader in RQMD than in the QMD. Between 6 fm/c and 9 fm/c,
the first collisions happen and thus particles start
to accumulate in  the mid-rapidity region. When one sees fig. 2, one expects
 that due to more collisions, QMD should show more
 stopping at 9 fm than RQMD.
Interestingly enough at 12 fm/c,
one finds just the reverse situation than at 9 fm/c. i.e.
  RQMD at 12 fm/c shows more
stopping. This is quite understandable when we remember that in the
time span between 9 fm/c and 15 fm/c, RQMD shows
a faster rise in the number of collisions as compared to the QMD
 and hence
more particles are stopped in RQMD than in QMD. After 12 fm/c
the reaction in RQMD is already nearly asymptotic, whereas
the reaction in the QMD still goes on.
As a result, the rapidity distribution in
RQMD at 12 fm/c and 18 fm/c is nearly the same whereas in the QMD particles
are still interacting and thus more and more particles are stopped
and hence at last QMD simulations dominate the mid-rapidity zone after
18 fm/c. Finally
the rapidity distribution is shown in fig.15 for different collisions
 involving $^{12}$C
to $^{28}$Si for hard EOS. It is evident from fig. 15 that though
the degree of stopping varies with the mass of the colliding nuclei
but, the influence of the relativistic effects ( i.e. less stopping using
RQMD than in the QMD ) is the same.
\\
\vspace*{0.5cm}\\
{\large\bf 4. Summary and Outlook : }\\

In this paper,  we have investigated the dynamical relativistic effects
which are originating from a full covariant treatment of the dynamics
of heavy ion reactions.
For this purpose, the  Quantum Molecular Dynamics
with its covariant extension ( Relativistic Quantum Molecular
Dynamics ) was used.
This generalization of QMD to a full covariant RQMD is based on the
Constraint Hamiltonian Dynamics. For a complete
understanding of the relativistic effects, five
different systems
$^{12}$C-$^{12}$C, $^{16}$O-$^{16}$O, $^{20}$Ne-$^{20}$Ne,
 $^{28}$Si-$^{28}$Si and
$^{40}$Ca-$^{40}$Ca were considered at bombarding energies of 50, 250,
500, 750, 1000, 1250, 1500, 1750 and 2000 MeV/nucl.. In addition, a
wide range of impact parameters between 0 fm( central collisions) to
10 fm (peripheral collisions) was also investigated.  In this study, we have
concentrated fully on the observables which are least affected by
the different model inputs. Some of these quantities are: The nucleon density
, collision history, formation of the resonance matter(which stands for
resonance production), the rapidity distribution which shows the
 thermalization
in heavy-ion collisions and also reflects the
stopping of the nuclear matter in heavy ion collisions.

We have shown that the final state rapidity distribution using RQMD and
QMD at 50 MeV/nucl. gives very close agreement. At higher energies,
the covariant
treatment of the dynamics affects not only
the maximal values of the nucleon density, but it also
affects the shape and size of the hot and compressed zone. The time
evolution of density using RQMD shows a faster rise and also a faster
decline as compared to QMD.  The rapid fall of the density in RQMD
gives us a hint that the covariant treatment produces less
thermalization and less stopping of nuclear matter at
 relativistic
energies.  It shows that in RQMD, the particles feel
some kind of repulsion which can be due to the Lorentz-contraction of
the initial coordinate space,
  the covariant formulation of the interactions and
also due to the mutli-times of the particles. In addition, the covariant
treatment of the collisions can also give  different results.
The number of collisions using RQMD are found to be larger  than for
QMD. These number of collisions in  RQMD are less than in QMD with an initial
Lorentz-contracted distribution.  This Lorentz-contraction in the
initial distribution in RQMD is  counterbalanced by the covariant treatment of
the
interactions. But when one switches off the interactions i.e. when
one simulates heavy-ion collisions in a cascade-mode, one finds that
the RQMD simulations show far more collisions than what one gets
with a Lorentz-contracted QMD.  The ratio of collisions which are
blocked (Pauli-blocking) relative to all attempted collisions at
50 MeV/nucl. is the same using
RQMD and QMD whereas at higher energies, the covariant
treatment of the collisions and of the Pauli-blocking (RQMD)
shows more collisions
which are blocked as compared to QMD.

Resonance matter especially delta production  is found
to be affected also by the covariant treatment of the dynamics. RQMD
 shows less delta production than QMD.  The
relativistic effect in delta production increases with the increase
in the bombarding energy. This indicates that the relativistic
effects can influence the subthreshold
 production of $K^+$~'s, $\bar p$~'s etc. where
the largest part of the production comes from  resonance matter.

 A detailed investigation
of the thermalization i.e.
the rapidity distribution shows that RQMD gives
less stopping than QMD.  This is quite understandable
. The rapid fall of the density indicates that
the particles in  RQMD are less stopped and thus it results in less
stopping of nuclear matter.

The main result of our present investigation is that relativistic
effects are not strongly dependent on the parameters of the
model. This means that both soft and hard  nuclear equation of states show
the same results. Further the relativistic effects can be
understood  when we look to the density evolution. The rapid
decrease of the density indicates that RQMD produce less
stopping than QMD.  In other words, the particles are kicked out from the hot
and dense zone which immediately indicates that RQMD should
give less resonance production, thermalization and nuclear stopping.
These results are the same for all energies and masses considered here.
In conclusion, our investigation of the relativistic effects in heavy-ion
collisions at SIS energies shows that the influence of the
relativistic effects at SIS energies is for the observables
considered here of a not too large importance . This may change if
subthreshold production of heavier particles like K$^+$~'s and
antiprotons are considered.
\\
\vspace*{2cm}
\\
The package UNISCO contains fully integrated the QMD code based on the
latest code of J\"org Aichelin and coworkers and our RQMD code. The
authors are thankful to Prof. J\"org Aichelin for providing us
with his new QMD code.

\newpage
{\large\bf Figure Captions : }\\
\vspace*{0.2cm}\\
\\
{\bf Fig. 1} The time evolution of the maximum density $\rho^{max}$. Here the
density is calculated in a central sphere with radius of 2 fm and the
maximum value of density reached anywhere in this sphere is taken. The
reaction under consideration is
$^{40}$Ca-$^{40}$Ca at the bombarding energy E$_{lab}$ = 1.5 GeV/nucl. and
at the impact parameter b = 2 fm. The upper and lower parts of the
figure show the results with hard and soft EOS's, respectively.
\vspace*{0.3cm}\\
\\
{\bf Fig. 2} The average collision rate (dN$_{coll}$/dt) for the reaction
$^{40}$Ca-$^{40}$Ca at impact parameter b = 1 fm and incident energy
 1.5 GeV/nucl.. This collision rate includes the Pauli-
blocking for each scattered pair of baryons.
 RQMD and QMD are shown by solid and dashed
histograms.
\vspace*{0.3cm}\\
\\
{\bf Fig. 3} The same as in fig.2, but for
$^{12}$C-$^{12}$C, $^{16}$O-$^{16}$O, $^{20}$Ne-$^{20}$Ne,
 $^{28}$Si-$^{28}$Si at incident energy = 1.5 GeV/nucl. and
at impact parameter b = 0.25 b$^{max}$. Here a hard EOS is used.
\vspace*{0.2cm}\\
{\bf Fig. 4} The same as in fig. 2 but with further decomposition of the total
collision rate into the elastic channel (dN$_{elas.}$/dt) and
the inelastic
channel dN$_{inel.}$/dt. Left and right parts of
the figure are calculated with the hard and the soft EOS's,
 respectively. The upper part represents the
elastic collisions whereas the lower part is for inelastic collisions.
\vspace{0.3cm}\\
\\
{\bf Fig. 5} The evolution of the total number of collisions
per nucleon as a function of the reaction time. Here the results are shown
for
RQMD, QMD and QMD with a Lorentz-contracted initial distribution
[labeled as QMD(cont.)] . The reaction under consideration is
$^{40}$Ca-$^{40}$Ca at 1.5 GeV/nucl. and using both hard and soft EOS's.
Here impact parameter is b = 2 fm.
\vspace*{0.2cm}\\
\\
{\bf Fig. 6} The time evolution of the total number of collisions
per nucleon  for the same reaction in fig.5 but in the cascade-mode.
The upper part shows the calculation at incident energy 1.5 GeV/nucl.,
 the lower part  at 2 GeV/nucl..
\vspace*{0.2cm}\\
\\
{\bf Fig. 7} The percentage of the collisions which are Pauli-blocked
due to lack of available free phase space as a function of the bombarding
energy. Here we simulate
$^{40}$Ca-$^{40}$Ca at an impact parameter b = 2 fm. The
displayed results are at 60 fm/c.
\vspace*{0.2cm}\\
\\
{\bf Fig. 8} The number of delta's obtained at 60 fm/c as a function
of the impact parameter in the simulation of
$^{40}$Ca-$^{40}$Ca at an incident energy of 1.5 GeV/nucl..
The upper and lower
part represent the results using the hard and the soft EOS's, respectively.
\vspace*{0.2cm}\\
\\
{\bf Fig. 9} The same as in fig. 8, but for the number of delta's as a
function of the total mass of the colliding nuclei.
The upper and lower part of the figure are calculated with the
hard and the soft EOS's.
For the explanation of the straight lines, see text.
\vspace*{0.2cm}\\
\\
{\bf Fig. 10} The same as in fig. 8, but for the delta's obtained at final
stage as a function of the bombarding energy. The impact parameter
is 2 fm.
\vspace*{0.2cm}\\
\\
{\bf Fig. 11} The rapidity distribution of the final stage of the reaction
of
$^{40}$Ca-$^{40}$Ca at impact parameter b = 2 fm. Note that the
rapidity  is given in units of  the beam rapidity. The upper left and
right parts are at energies 50 MeV/nucl. and 500 MeV/nucl.,
respectively. The lower left and right parts represent the
results for 1 GeV/nucl. and 1.5 GeV/nucl., respectively.
Here hard EOS was used.
\vspace*{0.2cm}\\
\\
{\bf Fig.12} The rapidity distribution dN/dY as a function
of Y$_{cm}$. The reaction is
$^{40}$Ca-$^{40}$Ca at 1.5 GeV/nucl. The solid and dashed curves are
the results of RQMD and QMD at an impact parameter b = 0 fm.
The solid and dashed histograms represent the results of RQMD and
QMD at an impact parameter b = 6 fm.
\vspace*{0.2cm}\\
\\
{\bf Fig. 13} The rapidity distribution dN/dY as a function of
Y$_{cm}$. The reaction is
$^{40}$Ca-$^{40}$Ca at the impact parameter b = 2 fm and at an incident
energy of 1.0 GeV/nucl. The upper part of the figure is
calculated with
a cascade-mode whereas the lower part represents the results in
a Vlasov approach.
\vspace*{0.2cm}\\
\\
{\bf Fig. 14} The rapidity distribution dN/dY as a function of
Y$_{cm}$ for the reaction of
$^{40}$Ca-$^{40}$Ca at an incident energy 1.5 GeV/nucl. and at
an impact parameter b = 0 fm. Here four different times t = 6, 9,
 12 and 18 fm/c are chosen. These results are calculated using
the hard EOS.
\vspace*{0.2cm}\\
\\
{\bf Fig. 15} The final rapidity distribution as a function
of Y$_{cm}$. Here four different systems
$^{12}$C-$^{12}$C, $^{16}$O-$^{16}$O, $^{20}$Ne-$^{20}$Ne,
 $^{28}$Si-$^{28}$Si are considered at the incident energy = 1.5 GeV/nucl. and
the impact parameter b = 0.25 b$^{max}$. The hard EOS is used.
\end{document}